\documentclass[10pt,letterpaper]{article}
\usepackage{amssymb}
\usepackage{amsmath}
\usepackage{graphicx}
\usepackage{bm}
\usepackage{cite}
\usepackage{fullpage} 

\begin{document}

\title{Modal and Polarization Qubits in Ti:LiNbO$_3$ Photonic Circuits for a Universal Quantum Logic
Gate}
\maketitle
\hyphenation{wave-guide wave-guides}

\begin{center}
\author{Mohammed~F.~Saleh,$^{\bf 1}$ Giovanni~Di~Giuseppe,$^{\bf 2,3}$Bahaa~E.~A.~Saleh,$^{\bf 1,2}$ and Malvin~Carl~Teich$^{\bf1,4,5}$}

\textit{$^1$Quantum Photonics Laboratory, Department of Electrical \& Computer Engineering,\\
Boston University, Boston, MA 02215, USA\\
$^2$Quantum Photonics Laboratory, College of Optics and Photonics (CREOL),\\
University of Central Florida, Orlando, FL 32816, USA\\
$^3$School of Science and Technology, Physics Division, University of Camerino,\\
62032 Camerino (MC), Italy\\
$^4$Department of Physics, Boston University, Boston, MA 02215, USA\\
$^5$Department of Electrical Engineering, Columbia University, New
York, NY 10027, USA}

\end{center}

\begin{abstract}
Lithium niobate photonic circuits have the salutary property of
permitting the generation, transmission, and processing of photons
to be accommodated on a single chip. Compact photonic circuits such
as these, with multiple components integrated on a single chip, are
crucial for efficiently implementing quantum information processing
schemes. We present a set of basic transformations that are useful
for manipulating modal qubits in Ti:LiNbO$_3$ photonic quantum
circuits. These include the mode analyzer, a device that separates
the even and odd components of a state into two separate spatial
paths; the mode rotator, which rotates the state by an angle in mode
space; and modal Pauli spin operators that effect related
operations. We also describe the design of a deterministic,
two-qubit, single-photon, CNOT gate, a key element in certain sets
of universal quantum logic gates. It is implemented as a
Ti:LiNbO$_3$ photonic quantum circuit in which the polarization and
mode number of a single photon serve as the control and target
qubits, respectively. It is shown that the effects of dispersion in
the CNOT circuit can be mitigated by augmenting it with an
additional path. The performance of all of these components are
confirmed by numerical simulations. The implementation of these
transformations relies on selective and controllable power coupling
among single- and two-mode waveguides, as well as the polarization
sensitivity of the Pockels coefficients in LiNbO$_3$.
\end{abstract}

\section{Introduction}

We recently investigated the possibility of using spontaneous
parametric down-conversion (SPDC) in two-mode waveguides to generate
guided-wave photon pairs entangled in mode number, using a cw pump
source. If one photon is generated in the fundamental (even) mode,
the other will be in the first-order (odd) mode, and \emph{vice
versa} \cite{Saleh09}. We also considered a number of detailed
photonic-circuit designs that make use of Ti:LiNbO$_{3}$ diffused
channel, two-mode waveguides for generating and separating photons
with various combinations of modal, spectral, and polarization
entanglement \cite{saleh10}. Selective mode coupling between
combinations of adjacent single-mode and two-mode waveguides is a
key feature of these circuits.

Although potassium titanyl phosphate (KTiOPO$_4$, KTP) single- and
multi-mode waveguide structures have also been used for producing
spontaneous parametric down-conversion
\cite{fiorentino07,Avenhaus09,Mosley09,zhong09,chen09}, it appears
that only the generation process, which makes use of a pulsed pump
source, has been incorporated on-chip. Substantial advances have
also recently been made in the development of single-mode
silica-on-silicon waveguide quantum circuits
\cite{Politi08,Matthews09}, with an eye toward quantum information
processing applications
\cite{Bennett98,Nielsen00,OBrien09,Politi09,Cincotti09,ladd10}. For
these materials, however, the photon-generation process necessarily
lies off-chip.

Lithium niobate photonic circuits have the distinct advantage that
they permit the \emph{generation}, \emph{transmission}, and
\emph{processing} of photons all to be achieved on a single chip
\cite{saleh10}. Moreover, lithium niobate offers a number of
ancillary advantages: 1) its properties are well-understood since it
is the basis of integrated-optics technology \cite{Nishihara89}; 2)
circuit elements, such as two-mode waveguides and
polarization-sensitive mode-separation structures, have low loss
\cite{saleh10}; 3) it exhibits an electro-optic effect that can
modify the refractive index at rates up to tens of GHz and is
polarization-sensitive \cite[Sec.~20.1D]{Saleh07}; and 4) periodic
poling of the second-order nonlinear optical coefficient is
straightforward so that phase-matched parametric interactions
\cite{Busacca04,Lee04}, such as SPDC and the generation of
entangled-photon pairs \cite{tanzilli01,dechatellus06}, can be
readily achieved. Moreover, consistency between simulation and
experimental measurement has been demonstrated in a whole host of
configurations
\cite{Alferness78,Alferness80,Hukriede03,Runde07,Runde08}. To
enhance tolerance to fabrication errors, photonic circuits can be
equipped with electro-optic adjustments. For example, an
electro-optically switched coupler with stepped phase-mismatch
reversal serves to maximize coupling between fabricated waveguides
\cite{schmidt76,kogelnik76}.

Compact photonic circuits with multiple components integrated on a
single chip, such as the ones considered here, are likely to be
highly important for the efficient implementation of devices in the
domain of quantum information science. The Controlled-NOT (CNOT)
gate is one such device. It plays an important role in quantum
information processing, in no small part because it is a key element
in certain sets of universal quantum logic gates (such as CNOT plus
rotation) that enable all operations possible on a quantum computer
to be executed \cite{Nielsen00,divincenzo95,knill01,ladd10}. Two
qubits are involved in its operation: a control and a target. The
CNOT gate functions by flipping the target qubit if and only if the
control qubit is in a particular state of the computational basis.
Two separate photons, or, alternatively, two different
degrees-of-freedom of the same photon, may be used for these two
qubits. A deterministic, two-qubit, single-photon, CNOT gate was
demonstrated using bulk optics in 2004 \cite{Fiorentino04}. More
recently, a probabilistic, two-photon, version of the CNOT gate was
implemented as a silica-on-silicon photonic quantum circuit; an
external bulk-optics source of polarization qubits was required,
however \cite{Politi08}. It is worthy of mention that qubit
decoherence is likely to be minimal in photonic quantum circuits;
however, decoherence resulting from loss in long waveguides can be
mitigated by the use of either a qubit amplifier \cite{gisin10} or
teleportation and error-correcting techniques \cite{Glancy04}.

This paper describes a set of basic building blocks useful for
manipulating modal qubits in Ti:LiNbO$_3$ photonic quantum circuits.
Section~2 provides a brief description of the geometry and
properties of the diffused channel Ti:LiNbO$_3$ waveguides used in
the simulations. Modal qubits are characterized in Sec.~3. Section~4
addresses the coupling of modes between two adjacent waveguides;
several special cases are highlighted. The principle of operation of
the mode analyzer, which separates the even and odd components of an
incoming state into two separate spatial paths, is set forth in
Sec.~5, as are the effects of the modal Pauli spin operator
$\sigma_z$. The mode rotator, which rotates the state by an angle in
mode space, is examined in Sec.~6, as is the modal Pauli spin
operator $\sigma_x$. Section~7 is devoted to describing the design
of a deterministic, two-qubit, single-photon, CNOT gate implemented
as a Ti:LiNbO$_3$ photonic quantum circuit, in which the
polarization and mode number of a single photon serve as the control
and target qubits, respectively. The conclusion is presented in
Sec.~8.

\section{Diffused channel Ti:LiNbO$_{3}$ waveguides}
\begin{figure}
\centering
\includegraphics[width=2 in,totalheight =1.2 in]{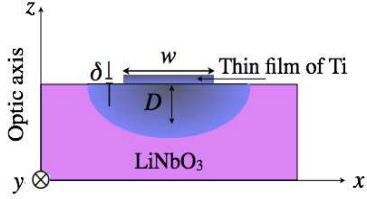}
\caption{Cross-sectional view of the fabrication of a diffused
channel Ti:LiNbO$_{3}$ waveguide (not to scale). A thin film of
titanium of thickness $\delta \approx 100$ nm and width $w$ is
diffused into a $z$-cut, $y$-propagating LiNbO$_3$ crystal. The
diffusion length $D = 3\: \mu$m.}  \label{ChannelWGs}
\end{figure}

All of the simulations presented in this paper refer to structures
that make use of Ti:LiNbO$_{3}$ diffused channel waveguides, as
illustrated in Fig.~\ref{ChannelWGs}. These waveguides are
fabricated by diffusing a thin film of titanium (Ti), with thickness
$ \delta \approx 100$ nm and width $w$, into a $z$-cut,
$y$-propagating LiNbO$_{3}$ crystal. The diffusion length $D$ is
taken to be the same in the two transverse directions: $D = 3\: \mu
$m. The TE mode polarized in the $x$-direction sees the ordinary
refractive index $n_{o}$, whereas the TM mode polarized in the
$z$-direction (along the optic axis) sees the extraordinary
refractive index $n_{e}$.

The ordinary and extraordinary refractive indices may be calculated
by making use of the Sellmeier equations \cite[Chap.~5]{Saleh07},
\cite{jundt97,Wong02}. The refractive-index increase introduced by
titanium indiffusion is characterized by $\Delta
n=2\delta\rho\,\,\mathrm{erf}\!\left( w/2D\right) /\sqrt{\pi}\,D $,
where $ \rho = 0.47$ and $0.625$ for $n_{o}$ and $n_{e}$,
respectively \cite{Feit83}. To accommodate wavelength dispersion,
$\Delta n$ can be modified by incorporating the weak factor $\xi =
0.052 + 0.065/\lambda^{2}$, where the wavelength $\lambda$ is
specified in $\mu$m \cite{Hutcheson87}. We calculate the effective
refractive index $n_{\mathrm{eff}}$ of a confined mode in two ways:
1) by using the effective-index method described in \cite{Hocker77};
and 2) by making use of the commercial photonic and network design
software package RSoft. The propagation constant of a guided mode is
related to $n_{\mathrm{eff}}$ via $\beta = 2\pi
n_{\mathrm{eff}}/\lambda$.

Applying a steady electric field to this structure in the
$z$-direction (along the optic axis) changes the ordinary and
extraordinary refractive indices of this uniaxial (trigonal 3$m$)
material by $\,-\frac12 n_{o}^{3}r_{13}V/d\,$ and $\,-\frac12
n_{e}^{3}r_{33}V/d\,$, respectively \cite[Example~20.2-1]{Saleh07},
where $V$ is the applied voltage; $d$ is the separation between the
electrodes; and $r_{13}$ and $r_{33}$ are the tensor elements of the
Pockels coefficient, which have values 10.9 and 32.6 pm/V,
respectively \cite{Wong02}.

\section{Modal qubits}

A qubit is a pure quantum state that resides in a two-dimensional
Hilbert space. It represents a coherent superposition of the basis
states, generally denoted $|0 \rangle$ and $|1 \rangle$. A qubit can
be encoded in any of several degrees-of-freedom of a single photon,
such as polarization \cite{bennett84}, spatial parity
\cite{abouraddy07}, or the mode number of a single photon confined
to a two-mode waveguide \cite{Saleh09,saleh10}. The Poincar{\'e}
sphere provides a geometrical representation for the state of a
modal qubit, much as it does for polarization
\cite[Sec.~6.1A]{Saleh07} and spatial parity \cite{yarnall07b}.

Indeed, polarization offers an intrinsically binary basis and is
often used to realize a qubit. However, the spatial modes of a
photon in a two-mode waveguide, one of which is even and the other
odd, are also binary and can therefore also be used to represent a
qubit. Modal qubits are particularly suited to photonic quantum
circuits since they can be both generated and easily transformed
on-chip by making use of elements such as mode analyzers, mode
rotators, and two-mode electro-optic directional couplers. The modal
space of a two-mode waveguide therefore offers an appealing
alternative to polarization for representing qubits in quantum
photonic circuits.

The comparison between modal and spatial-parity qubits is
instructive. Spatial-parity qubits are defined on a 2D Hilbert space
in which the 1D transverse spatial modes of the photon are
decomposed into even and odd spatial-parity components
\cite{abouraddy07,yarnall07b,yarnall07a}. Modal qubits also relate
to parity, but in a simpler way. They are defined on a 2D Hilbert
space in which the bases are a single 1D even-parity function and a
single 1D odd-parity function. These two functions are the
fundamental (even, $m=0$) and first-order order (odd, $m=1$)
transverse spatial eigenmodes of the Helmholtz equation for a
two-mode waveguide.

Photon pairs can be exploited for use in quantum photonic circuits
\cite{Politi08,saleh10}, as well as for producing heralded
single-photon pure states \cite{saleh85} in well-defined
spatiotemporal modes, which are required for many quantum
information technology applications such as quantum cryptography
\cite{Gisin02} and linear optical quantum computing \cite{knill01}.
Care must be taken, however, to ensure that the intrinsic quantum
correlations between the twin photons are eliminated so that the
surviving photon is in a pure state
\cite{abouraddy01,mosley08,levine10}. One way of achieving this is
to generate the twin photons with a factorable joint amplitude
\cite{grice01,walton03,walton04,carrasco04}. We have previously
shown that a Type-0 interaction could be used to generate photon
pairs that are degenerate in frequency and polarization, but with
opposite mode number \cite[Sec.~3]{saleh10}. Coupling these photons
into two single-mode waveguides would allow one of these photons to
be used to herald the arrival of the other. The heralded photon
could then be coupled into a two-mode waveguide which, with the
addition of a mode rotator, would serve as a source of modal qubits.
Such a source would be analogous to the one fashioned from bulk
optics by Fiorentino et al. \cite{Fiorentino04} using Type-II SPDC.
However, the Type-0 source of modal qubits described above would be
on-chip and would also make use of the strongest nonlinear component
of the second-order tensor, $d_{33}$, thereby enhancing the
efficiency of the interaction \cite{Myers95}.

The quantum state of a single photon in a two-mode waveguide,
assuming that its polarization is TE or TM, can be expressed as
$\vert\Psi\rangle =\alpha_{\,1}\vert e\rangle+\alpha_{\,2}\vert
o\rangle$, where $\vert e\rangle$ and $\vert o\rangle$ represent the
even and odd basis states, respectively;
and $\alpha_{\,1}$ and $\alpha_{\,2}$ are their weights. All
operations on the single-photon state are effected via auxiliary
adjacent waveguides, which are sometimes single-mode and sometimes
two-mode. We exploit the concepts of selective and controllable
coupling between waveguides, together with the isomorphism between
waveguide coupling and the SO(2) rotation matrix, to design a mode
analyzer, a mode rotator, modal Pauli spin operators, and a CNOT
gate useful for quantum information processing.

\section{Mode coupling between adjacent waveguides}
The coupling between two lossless, single-mode waveguides is
described by a unitary matrix $\mathbf{T}$ that takes the form
\cite[Sec.~8.5B]{Saleh07}
\begin{equation}
\mathbf{T}=\left[ \begin{array}{cc}
 A & -jB  \\[1mm]
-jB^{*} & A^{*}
\end{array}\right] ,
\end{equation}
where \,\,$A= \exp\left(j\Delta\beta \,L/2\right) \left[\cos \gamma
L-j({\Delta\beta}/{2\gamma})\,\sin\gamma L\right]$\,\,\, and \, $B=
({\kappa}/{\gamma})\, \exp\left(j\Delta\beta \,L/2\right)
\,\sin\gamma L\,$. Here, $\Delta \beta$ is the phase mismatch per
unit length between the two coupled modes; $L$ is the coupling
interaction length; $\kappa$ is the coupling coefficient, which
depends on the widths of the waveguides and their separation as well as on the mode profiles; $\gamma^{2} = \kappa^{2}+\frac14\Delta \beta^{2}$; and the
symbol $^*$ represents complex conjugation.

This unitary matrix $\mathbf{T}$ can equivalently be written in
polar notation as \cite{Buhl87}
\begin{equation}
\mathbf{T}=\left[ \begin{array}{cc}
\cos\left( \theta/2\right) \,\exp\left( j\phi_{A}\right)
& -j\sin\left( \theta/2\right) \,\exp\left( j\phi_{B}\right) \\[1mm]
-j\sin\left( \theta/2\right) \,\exp\left(
-j\phi_{B}\right)&\cos\left( \theta/2\right) \,\exp\left(
-j\phi_{A}\right)
\end{array}\right],
\label{eq:DCmatrix}
\end{equation}
where \;\; $\theta=2\sin^{-1}\left[ ({\kappa}/{\gamma})\sin\gamma L
\right]$; \;\;$\phi_{A}= \phi_{B}+\tan^{-1}\left[({-\Delta\beta
}/{2\gamma}) \tan\gamma L  \right]$; \;\; and \;\;$\phi_{B}=
{\Delta\beta L}/{2}$. Using this representation, the coupling
between the two waveguides can be regarded as a cascade of three
processes: 1) phase retardation, 2) rotation, and 3) phase
retardation. This becomes apparent if Eq.~(\ref{eq:DCmatrix}) is
rewritten as
\begin{equation}\label{eq:cascade}
\mathbf{T}=\exp\left( -j\phi_{B}\right)
\mathbf{T}_{3}\,\mathbf{T}_{2}\,\mathbf{T}_{1}\,,
\end{equation}
with
\begin{equation}\label{eq:psrotps}
\mathbf{T}_{1}=\left[ \begin{array}{cc}
 1& 0 \\[1mm]
0& e^{-j\Gamma_{1}}
\end{array}\right];\;\;\;\;
\mathbf{T}_{2}=\left[ \begin{array}{cc}
\cos\left( \theta/2\right)  & -j\sin\left( \theta/2\right) \\[1mm]
-j\sin\left( \theta/2\right)&\cos\left( \theta/2\right)
\end{array}\right];\;\;\;\;
\mathbf{T}_{3}=\left[ \begin{array}{cc}
 e^{-j\Gamma_{2}} & 0 \\[1mm]
0&  1
\end{array}\right]\,,
\end{equation}
where $\Gamma_{1}= \phi_{A}-\phi_{B}$;\; $\Gamma_{2}=
-\phi_{A}-\phi_{B}$;\; and $\mathbf{T}_{1}$, $\mathbf{T}_{2}$, and
$\mathbf{T}_{3}$ represent, in consecutive order, phase retardation,
rotation, and phase retardation. The phase shift $\phi_B$ is a
constant of no consequence.

For perfect phase matching between the coupled modes, i.e., for
$\Delta\beta =0$ and an interaction coupling length
$L=q\pi/2\kappa$, where $q$ is an odd positive integer, the coupling
matrix $\mathbf{T}$ reduces to
\begin{equation} \label{eq:TforZeroDeltaBeta}
\mathbf{T}=\exp\!\left(\dfrac{ jq\pi}{2}\right)\left[
\begin{array}{cc}
 0 & -1  \\[1mm]
-1 & 0
\end{array}\right],
\end{equation}
indicating that the modes are flipped. Applying this operation twice
serves to double flip the vector, thereby reproducing the input, but
with a phase shift twice that of $q\pi/2.$ On the other hand, for
$\gamma L =p\pi$, with $p$ an integer, the matrix becomes
\begin{equation}
\mathbf{T}=(-1)^p \left[ \begin{array}{cc}
 \exp\left( j\phi_{A}\right) & 0 \\[1mm]
0& \exp\left( -j\phi_{A}\right)
\end{array}\right].
\end{equation}
Finally, for weak coupling $\left(\kappa \approx 0 \; \mathrm{or} \;
\kappa \ll \Delta\beta\right)$, we have $\phi_{A}\approx 0$,
whereupon $\mathbf{T}$ reduces to the identity matrix.

Our interest is in three scenarios: 1) coupling between a pair of
single-mode waveguides (SMWs); 2) coupling between a pair of
two-mode waveguides (TMWs); and 3) coupling between a SMW and a TMW.
The matrix described in Eq.~(\ref{eq:DCmatrix}) is not adequate for
describing the coupling in the latter two cases; in general, a
$4\times 4$ matrix is clearly required for describing the coupling
between two TMWs. However, for the particular cases of interest
here, the coupling between the two waveguides is such that only a
single mode in each waveguide participates; this is because the
phase-matching conditions between the interacting modes are either
satisfied --- or not satisfied. As an example for identical
waveguides, similar modes couple whereas dissimilar modes fail to
couple as a result of the large phase mismatch. The net result is
that, for the cases at hand, the general matrix described in
Eq.~(\ref{eq:DCmatrix}) reduces to submatrices of size $2\times 2$,
each characterizing the coupling between a pair of modes.

\section{Mode analyzer and modal {P}auli spin operator $\sigma_z$}

\begin{figure}
\centering
\includegraphics[width=5in,totalheight =1.6 in]{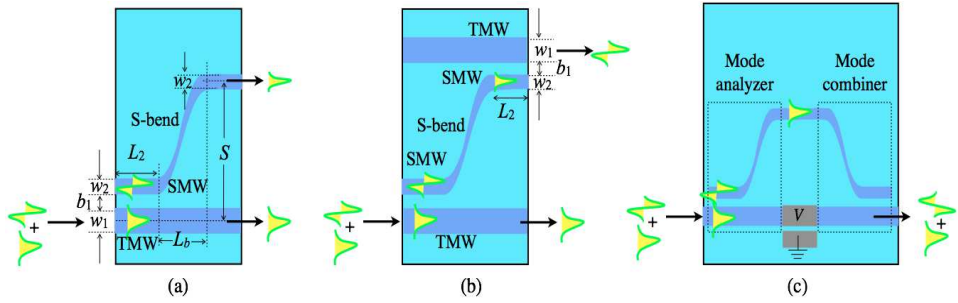}
\caption{(a) Sketch of a photonic circuit that serves as a mode
analyzer (not to scale). It is implemented by bringing a single-mode
waveguide (SMW) of width $w_{2}$ and length $L_{2}$ into proximity
with a two-mode waveguide (TMW) of width $w_{1}$. The two waveguides
are separated by a distance $b_1$. An $S$-bend waveguide of initial
and final width $w_{2}$, and bending length $L_{b}$, is attached to
the end of the SMW. The center-to-center separation between the
output of the $S$-bend and the TMW is denoted $S$. All $S$-bends
considered in this paper have dimensions $L_{b} = 10$ mm and $S =
127 \:\mu$m (the standard spatial separation \cite{Runde08}). The
odd mode is separated and delivered as an even distribution. (b)
Sketch of a mode analyzer (not to scale) that separates the odd mode
and delivers it as an odd distribution. It is more complex than the
design presented in (a) because it incorporates a second TMW, again
of width $w_{1}$, that is brought into proximity with a SMW of width
$w_{2}$ and length $L_{2}$ placed at the output of the $S$-bend.
These two waveguides are again separated by a distance $b_1$. (c)
Sketch of a photonic circuit (not to scale) that changes the sign of
the odd mode while leaving the even mode intact, thereby
implementing the modal Pauli spin operator $\sigma_z$. An
electro-optic phase modulator is used to compensate for any
unintended differences in the phase delays encountered by the even
and odd modes as they transit the circuit.} \label{Modeanalyzer}
\end{figure}

A \emph{mode analyzer} is a device that separates the even and odd
components of an incoming state into two separate spatial paths. It
is similar to the \emph{parity analyzer} of one-photon parity space
\cite{abouraddy07}. For the problem at hand, its operating principle
is based on the selective coupling between adjacent waveguides of
different widths. The even and odd modes of a TMW of width $w_{1}$
are characterized by different propagation constants. An auxiliary
SMW (with appropriate width $w_{2}$, length $L_{2}$, and separation
distance $b_1$ from the TMW) can be used to extract only the odd
component \cite{saleh10}. The result is a mode analyzer that
separates the components of the incoming state, delivering the the
odd mode as an even distribution, as shown in
Fig.~\ref{Modeanalyzer}(a). The end of the SMW is attached to an
$S$-bend waveguide, with initial and final widths $w_{2}$, to
obviate the possibility of further unwanted coupling to the TMW and
to provide a well-separated output port for the extracted mode. If
it is desired that the output be delivered as an odd distribution
instead, another SMW to TMW coupling region (with the same
parameters) may be arranged at the output end of the $S$-bend, as
illustrated in Fig.~\ref{Modeanalyzer}(b). This allows the
propagating even mode in the SMW to couple to the odd mode of the
second TMW, thereby delivering an odd distribution at the output.
The appropriate coupler configuration is determined by the
application at hand. It is important to note that the mode analyzer
is a bidirectional device: it can be regarded as a \emph{mode
combiner} when operated in the reverse direction, as we will soon
see.

The Pauli spin (or spatial-parity) operator $\sigma_{z}$ introduces
a phase shift of $\pi$ (imparts a negative sign) to the odd
component of the photon state, leaving the even component unchanged;
it thus acts as a half-wave retarder in mode space. It can be
implemented by exploiting modal dispersion between the even and odd
modes: a single TMW of length
$\pi/\left|\beta_{e}-\beta_{o}\right|$, where $\beta_{e}$ and
$\beta_{o}$ are the propagation constants of the even and odd modes,
respectively, results in the desired phase shift of $\pi$. For a
weakly dispersive medium, however,  a waveguide longer than
practicable might be required. An alternative approach for
implementing the Pauli spin operator $\sigma_{z}$ involves cascading
a mode analyzer and a mode combiner, as illustrated in
Fig.~\ref{Modeanalyzer}(c). As established in
Eq.~(\ref{eq:TforZeroDeltaBeta}), perfect coupling between a pair of
adjacent waveguides over an interaction length $L = q \pi /2\kappa$
introduces a phase shift of $q \pi /2$, where $q$ is an odd positive
integer. A cascade of two such couplings thus results in a phase
shift $q \pi$, with $q$ odd, thereby implementing the Pauli spin
operator $\sigma_{z}$. Proper design dictates that
$\beta_{e}L_{e}=\beta_{o}L_{o}$, where $L_{e}$ and $L_{o}$ are the
distances traveled by the even and odd modes, respectively.
Imperfections in the fabrication of the circuit may be compensated
by making use of an electro-optic (EO) phase modulator, as sketched
in Fig.~\ref{Modeanalyzer}(c).

\begin{figure}
\centering
\includegraphics[width=4 in,totalheight =1.5 in]{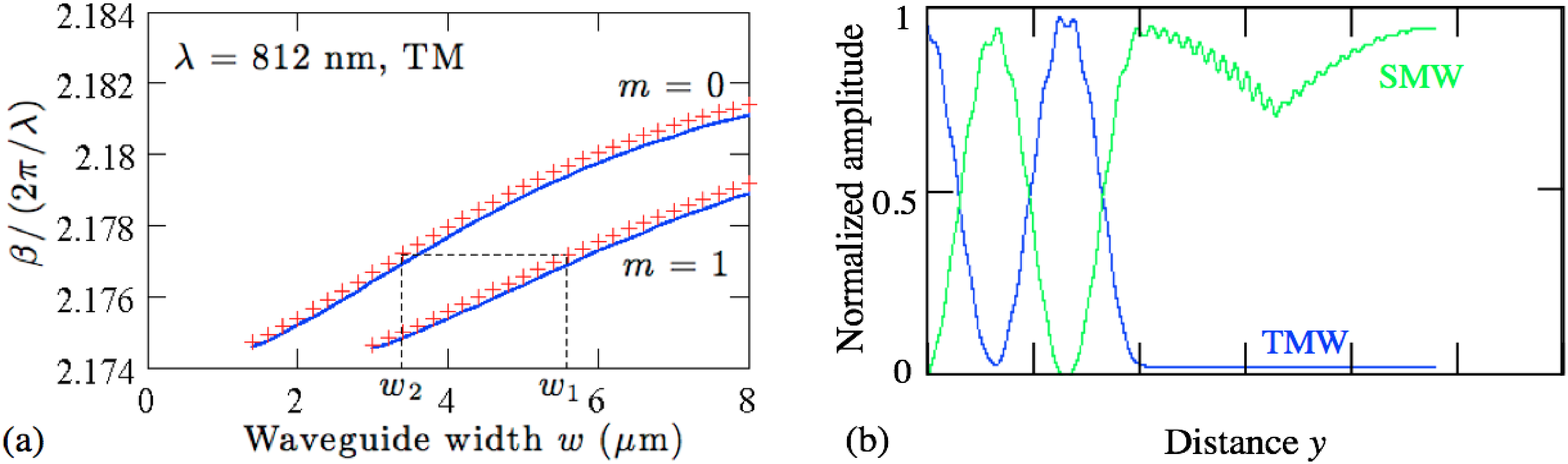}
\caption{(a) Dependencies of the normalized propagation constants
$\beta$ of the fundamental ($m=0$) and first-order ($m=1$) modes on
the widths $w$ of the diffused channel Ti:LiNbO$_{3}$ waveguides.
The input wave has wavelength $\lambda = 0.812 \:\mu$m and TM
polarization. The solid curves were obtained using the
effective-index method described in \cite{Hocker77}, whereas the
plus signs were computed using the software package RSoft. The
dotted vertical lines represent the desired widths $w_1$ and $w_2$.
(b) Simulated performance of a mode analyzer that takes the form
displayed in Fig.~\ref{Modeanalyzer}(a). The blue curve represents
the evolution with distance of the normalized amplitude of the odd
mode in a TMW of width $w_{1} = 5.6 \:\mu$m, whereas the green curve
shows the evolution of the even mode in a SMW of dimensions $w_{2} =
3.4 \:\mu$m and $L_{2} = 6.2$ mm. The separation between the TMW and
the SMW is  $b_1 = 4 \:\mu$m and the $S$-bend has dimensions $L_{b}
= 10$ mm and $S = 127 \:\mu$m. The dip in the curve for the SMW is
associated with the tapered nature of the $S$-bend. The results were
obtained with the help of the software package RSoft.}
\label{Linearcoupling}
\end{figure}

An example illustrating the operation of a mode analyzer, such as
that shown in Fig.~\ref{Modeanalyzer}(a), is provided in
Fig.~\ref{Linearcoupling}. The behavior of the normalized
propagation constants $\mathrm{\beta}$ of the even ($m=0$) and odd
($m=1$) modes before Ti indiffusion, as a function of the waveguide
width $w$, is presented in Fig.~\ref{Linearcoupling}(a) for TM
polarization at a wavelength of $\lambda = 0.812 \:\mu$m. The
horizontal dotted line crossing the two curves represents the
phase-matching condition for an even and an odd mode in two
waveguides of different widths. The simulation presented in
Fig.~\ref{Linearcoupling}(b) displays the evolution of the
normalized amplitudes of the two interacting modes with distance.

\section{Mode rotator and modal {P}auli spin operator $\sigma_x$}

The \emph{mode rotator} is an operator that rotates the state by an
angle $\theta$ in mode space, just as a polarization rotator rotates
the polarization state. It is also analogous to the \emph{parity
rotator} of one-photon spatial-parity space \cite{abouraddy07}. It
achieves rotation by cascading a mode analyzer, a \emph{directional
coupler}, and a mode combiner; the three devices are regulated by
separate EO phase modulators to which external voltages are applied.
The mode analyzer splits the incoming one-photon state into its even
and odd projections; the directional coupler mixes them; and the
mode combiner recombines them into a single output.

Implementation of the mode rotator is simplified by making use of
the factorization property of the unitary matrix $\mathbf{T}$ that
characterizes mode coupling in two adjacent waveguides (see Sec.~4).
As shown in Eqs.~(\ref{eq:cascade}) and (\ref{eq:psrotps}), the
coupling between two lossless waveguides can be regarded as a
cascade of three stages: phase retardation, rotation, and phase
retardation. If the phase-retardation components were eliminated,
only pure rotation, characterized by the SO(2) operator, would
remain.

The phase-retardation components can indeed be compensated by making
use of a pair of EO phase modulators to introduce phase shifts of
$\Gamma_{1}$ and $\Gamma_{2}$, before and after the EO directional
coupler, respectively. These simple U(1) transformations convert
$\mathbf{T}_{1}$ and $\mathbf{T}_{3}$ in Eq.~(\ref{eq:psrotps}) into
identity matrices, whereupon Eq.~(\ref{eq:cascade}) becomes the
SO(2) rotation operator. For a mode of wavelength $\lambda$, and an
EO phase modulator of length $L$ and distance $d$ between the
electrodes, the voltage required to introduce a phase shift of
$\Gamma$ is $V=\lambda\,d\,\Gamma/\pi\,r\,n^{3}L\,$, where the
Pockels coefficient $r$ assumes the values $r_{13}$ and $r_{33}$,
for $n= n_{o}$ and $n= n_{e}$, respectively
\cite[Sec.~20.1B]{Saleh07}.

\begin{figure}
\centering
\includegraphics[width=3.2 in,totalheight =1.2 in]{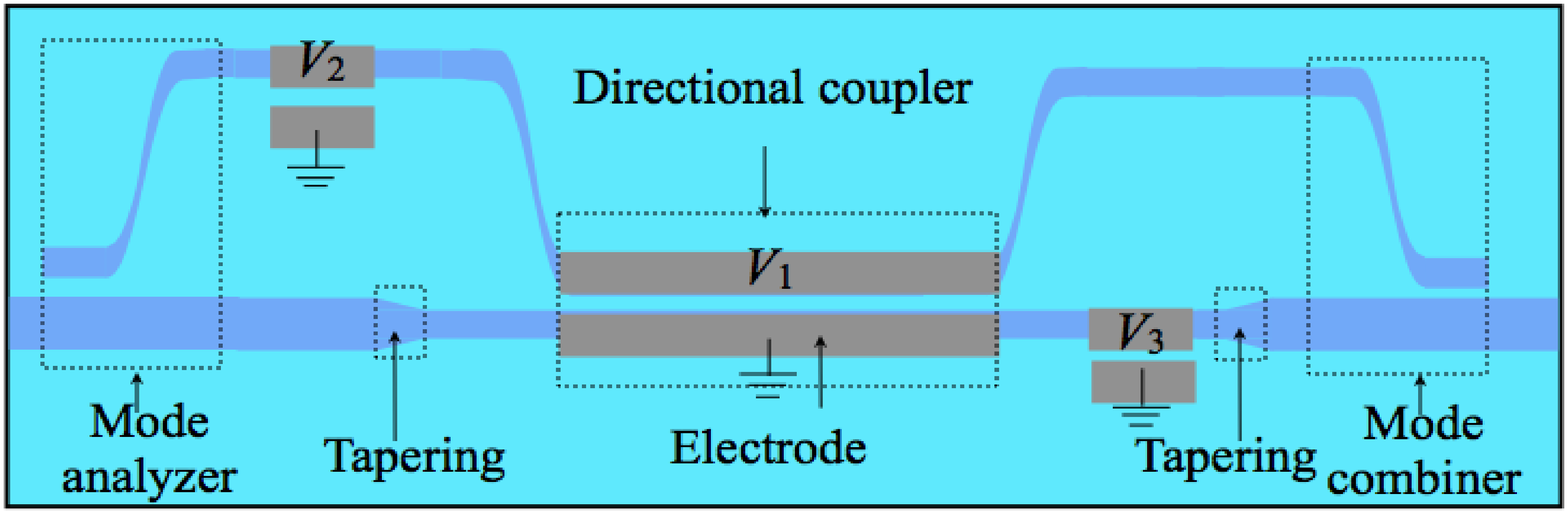}
\caption{Sketch of a photonic circuit that serves as a mode rotator
(not to scale). It is implemented by sandwiching a directional
coupler between a mode analyzer and a mode combiner. The coupling
length of the directional coupler is $\pi/2\kappa$. To obtain a
specified angle of rotation $\theta$, voltages $V_{1}$, $V_{2}$, and
$V_{3}$ are applied to the EO directional coupler, the input EO
phase modulator, and the output EO phase modulator, respectively.}
\label{Moderotator}
\end{figure}
The standard EO directional coupler consists of two adjacent
identical SMWs and makes use of an EO phase modulator to control the
transfer of modal power between them \cite[Sec.~20.1D]{Saleh07}.
When no voltage is applied to the EO modulator, the optical power is
totally transferred from one waveguide to the other, provided that
the interaction length $L$ over which they interact is an odd
integer multiple of the coupling length, $\pi/2\kappa$
\cite[Sec.~8.5B]{Saleh07}. The application of a voltage to the EO
modulator introduces a phase mismatch between the two interacting
modes that results in partial, rather than full, optical power
transfer. In particular, if the voltage is chosen such that
$|\Delta\beta L \,| = \sqrt{3}\pi$ (or
$\sqrt{7}\pi,\sqrt{11}\pi,\ldots)$, then no power is transferred
between the two waveguides. The voltage required to introduce a
phase mismatch of $\Delta \beta$ is approximately
$V=\lambda\,d\,\Delta\beta/2\pi\,r\,n^{3}\,$
\cite[Sec.~20.1D]{Saleh07}. The waveguide beam combiner suggested by
Buhl and Alferness \cite{Buhl87} operates on the same principle.

However, because our modal state resides in a TMW, rather than in a
SMW associated with the usual directional coupler, a mode analyzer
with a configuration similar to that shown in
Fig.~\ref{Modeanalyzer}(a) is used to direct the odd component to
one arm of the EO directional coupler, and the even component to the
other arm through an adiabatically tapered region, as shown in
Fig.~\ref{Moderotator}. A mirror-image tapered region and mode
combiner follow the directional coupler to recombine the two
components at the output of the device. Voltages $V_{1}$, $V_{2}$,
and $V_{3}$ are applied to the EO directional coupler, the input EO
phase modulator, and the output EO phase modulator, respectively.
The voltages $V_{2}$ and $V_{3}$ can be modified as necessary to
ensure that the overall phases acquired by the odd and even modes,
both before and after the directional coupler, are identical when
$V_{1}=0$.

\begin{figure}
\centering
\includegraphics[width=3.2in,totalheight=2in]{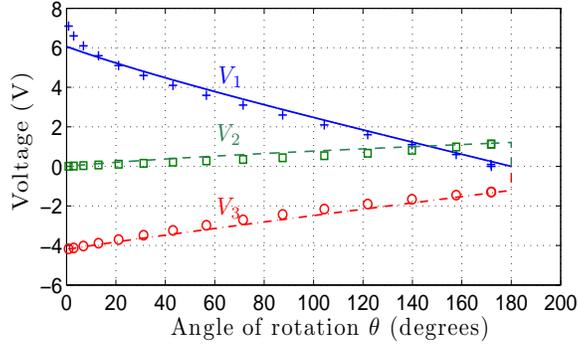}
\caption{Operating voltages for the mode rotator vs. the angle of
rotation $\theta$. Voltages $V_{1}$ (solid blue curve), $V_{2}$
(dashed green curve), and $V_{3}$ (dashed-dotted red curve) are
applied to the EO directional coupler, the input EO phase modulator,
and the output EO phase modulator, respectively. The input has
wavelength $\lambda = 0.812 \:\mu$m and TM polarization. The
directional coupler comprises two identical SMWs separated by $d = 5
\:\mu$m; each SMW has width $2.2 \:\mu$m and length $1.73$ mm. The
input and output EO phase modulators have electrode lengths of $5$
mm and electrode separations of $5 \:\mu$m. The curves represent
theoretical calculations while the symbols represent simulated data
obtained using the RSoft program.} \label{Deviceoperation}
\end{figure}
An example showing the operating voltages $V_{1}, V_{2}$, and
$V_{3}$ required to obtain a specified angle of rotation $\theta$ is
provided in Fig.~\ref{Deviceoperation}. The directional-coupler
voltage $V_{1}$ has an initial value (for $\theta = 0$) that
corresponds to a phase mismatch $|\Delta\beta L\,|=\sqrt{3}\,\pi$;
decreasing $V_{1}$ results in increasing $\theta$. When $V_{1} = 0$,
the angle of rotation is $\pi$; the device then acts as the Pauli
spin operator $\sigma_x$\,, which is a \emph{mode flipper}
(analogous to the \emph{parity flipper}
\cite{abouraddy07,yarnall07b}). For $V_{1} = 0$, there are an
infinite number of solutions for the values of $V_{2}$ and $V_{3}$,
provided, however, that $V_{2} = -V_{3}$.

\section{Controlled-NOT (CNOT) gate}

Deterministic quantum computation that involves several
degrees-of-freedom of a single photon for encoding multiple qubits
is not scalable inasmuch as it requires resources that grow
exponentially \cite{Fiorentino04}. Nevertheless, few-qubit quantum
processing can be implemented by exploiting multiple-qubit encoding
on single photons \cite{mitsumori03}. We propose a novel
\emph{deterministic, two-qubit, single-photon, CNOT gate},
implemented as a Ti:LiNbO$_3$ photonic quantum circuit, in which the
polarization and mode number of a single photon serve as the control
and target qubits, respectively.

\begin{figure}
\centering
\includegraphics[width=3.2 in,totalheight =1.4 in]{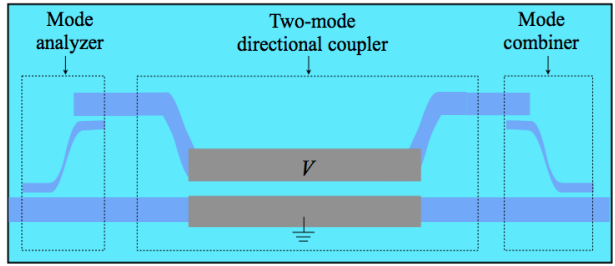}
\caption{Sketch of a Ti:LiNbO$_3$ photonic quantum circuit that
behaves as a novel deterministic, two-qubit, single-photon, CNOT
gate (not to scale). The control qubit is polarization and the
target qubit is mode number. The circuit bears some similarity to
the mode rotator shown in Fig.~\ref{Moderotator}; both are
implemented by sandwiching an EO directional coupler between a mode
analyzer and a mode combiner. However, for the CNOT gate, the EO
directional coupler comprises a pair of TMWs, whereas the mode
rotator uses a conventional EO directional coupler utilizing a pair
of SMWs.} \label{CNOT}
\end{figure}

The operation of this gate is implemented via a
\emph{polarization-sensitive, two-mode, electro-optic directional
coupler}, comprising a pair of identical TMWs integrated with an
electro-optic phase modulator, and sandwiched between a mode
analyzer and a mode combiner. It relies on the polarization
sensitivity of the Pockels coefficients in LiNbO$_3$. A sketch of
the circuit is provided in Fig.~\ref{CNOT}. The mode analyzer
spatially separates the even and odd components of the state for a
TM-polarized photon, sending the even component to one of the TMWs
and the odd component to the other. At a certain value of the EO
phase-modulator voltage, as explained below, the even and odd modes
can exchange power. The modified even and odd components are then
brought together by the mode combiner.

\begin{figure}
\centering
\includegraphics[width=3.2in,totalheight=2in]{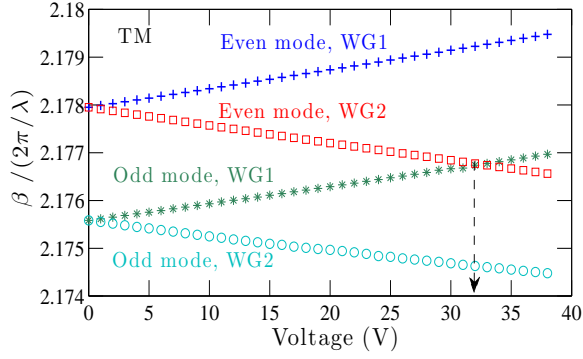}
\caption{Dependencies of the normalized propagation constants
$\beta$ on the voltage applied to an EO TMW directional coupler
comprising two waveguides [WG1 and WG2]. The propagation constants
differ for the even and odd modes except at one particular voltage
(vertical dashed line) where the even mode in one waveguide can be
phase-matched to the odd mode in the other waveguide. The TMWs are
identical, each of width $4 \:\mu $m, and they are separated by $4
\:\mu $m. The input has wavelength $\lambda = 0.812 \:\mu$m and TM
polarization. The symbols represent simulated data obtained using
the RSoft program.} \label{Exmodeconverter}
\end{figure}

To show that the device portrayed in Fig.~\ref{CNOT} operates as a
CNOT gate, we first demonstrate that the target qubit is indeed
flipped by a TM-polarized control qubit, so that $\vert 1\rangle
\equiv \vert {\rm TM}\rangle$. The polarization sensitivity of the
Ti:LiNbO$_3$ TMWs resides in the values of their refractive indices
$n$, which depend on the polarizations of the incident waves and the
voltage applied to its EO phase modulator; and on their Pockels
coefficients $r$, which depend on the polarization
\cite[Example~20.2-1]{Saleh07}. For a photon with TM polarization,
the two-mode EO directional coupler offers two operating regions
with markedly different properties. At low (or no) applied voltage,
interaction and power transfer take place only between like-parity
modes in the two waveguides because the propagation constants of the
even and odd modes are different, so they are not phase-matched.
However, at a particular higher value of the applied voltage, the
behavior of the device changes in such a way that only the even mode
in one waveguide, and the odd mode in the other, can interact and
exchange power. This arises because the refractive indices of the
two waveguides depend on the voltage applied to the device; they
move in opposite directions as the voltage increases since the
electric-field lines go downward in one waveguide and upward in the
other. Figure~\ref{Exmodeconverter} provides an example illustrating
the dependencies of the propagation constants of the even and odd
modes, in the two TMWs, as a function of the applied voltage.

At a voltage indicated by the vertical dashed line in
Fig.~\ref{Exmodeconverter}, the even mode in one waveguide is
phase-matched to the odd mode in the other. In a directional coupler
with suitable parameters, a TM-polarized control bit will then
result in a flip of the modal bit, whereupon $\alpha_{\,1}\vert
e\rangle+\alpha_{\,2}\vert o\rangle \rightarrow \alpha_{\,1}\vert
o\rangle+\alpha_{\,2}\vert e\rangle$. A TE-polarized control qubit,
on the other hand, which sees $n_o$ rather than $n_e$, will leave
the target qubit unchanged because of phase mismatch, so that $\vert
0\rangle \equiv \vert {\rm TE}\rangle$. Hence, the target qubit is
flipped if and only if the control qubit is $\vert 1\rangle$, and is
left unchanged if the control qubit is $\vert 0\rangle$, so that the
device portrayed in Fig.~\ref{CNOT} does indeed behave as a CNOT
gate. In principle, it would also be possible to use a TE-polarized
control qubit to flip the target bit; this option was not selected
because it would require a higher value of EO phase-modulator
voltage since the TE Pockels coefficient $r_{13}$ is smaller than
the TM Pockels coefficient $r_{33}$ \cite{Wong02}.

A drawback of the photonic circuit illustrated in Fig.~\ref{CNOT} is
that it suffers from the effects of dispersion, which is deleterious
to the operation of circuits used for many quantum information
applications. Dispersion results from the dependence of the
propagation constant $\beta$ on frequency, mode number, and
polarization. Polarization-mode dispersion generally outweighs the
other contributions, especially in a birefringent material such as
LiNbO$_{3}$.

\begin{figure}
\centering
\includegraphics[width=3.8 in,totalheight =2 in]{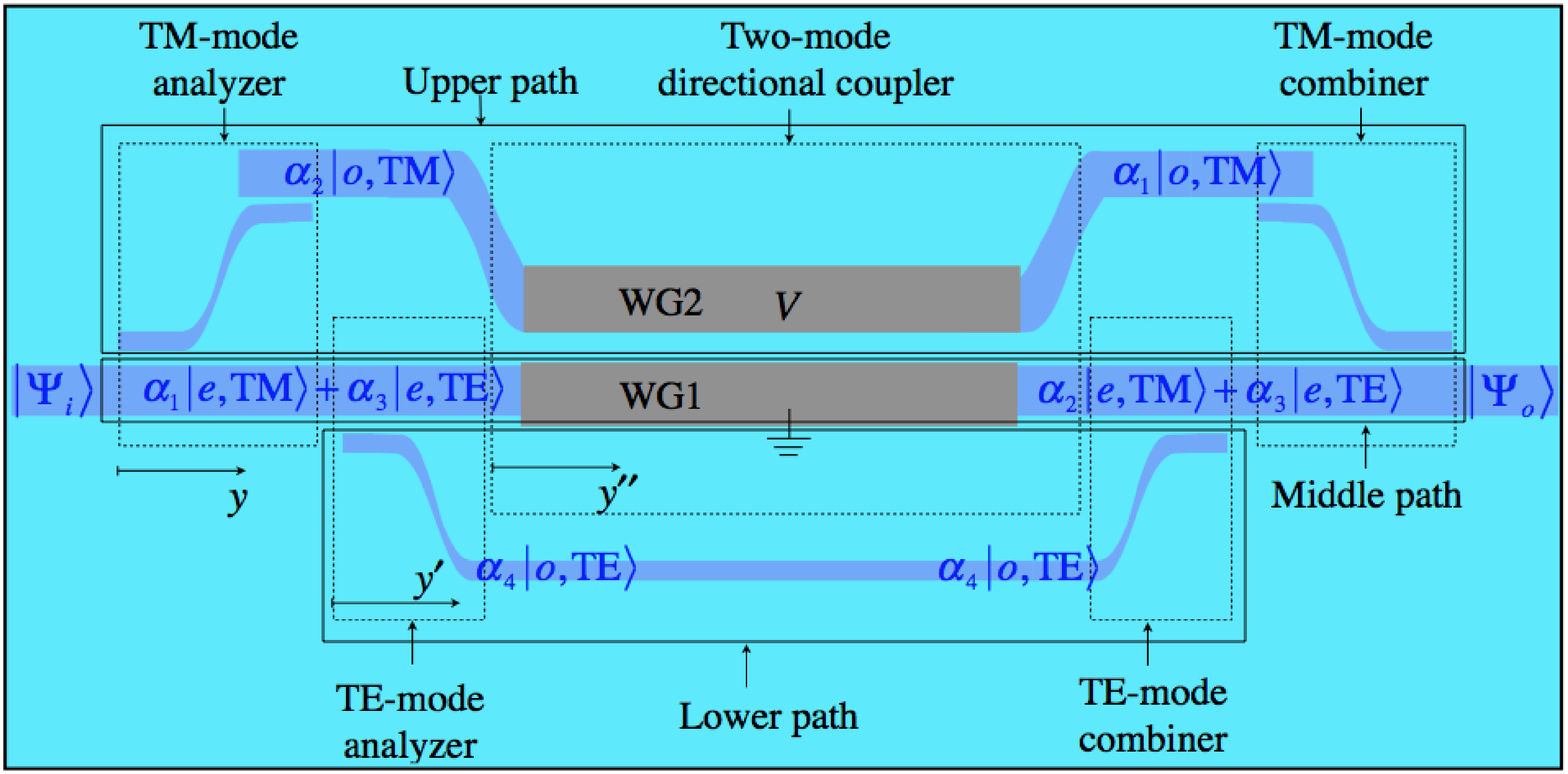}
\caption{Sketch of a Ti:LiNbO$_3$ photonic quantum circuit that
behaves as a novel dispersion-managed, deterministic, two-qubit,
single-photon, CNOT gate (not to scale). The control qubit is
polarization and the target qubit is mode number. The design is more
complex than that shown in Fig.~\ref{CNOT} because it accommodates
dispersion management via path-length adjustments of the upper,
middle, and lower paths. An EO TMW directional coupler is sandwiched
between polarization-sensitive mode analyzers and
polarization-sensitive mode combiners. The lower and upper
waveguides of the two-mode directional coupler are denoted WG1 and
WG2, respectively. The paths taken by the components of the input
state $\vert\Psi_{i}\rangle$ are shown, as is the output state
$\vert\Psi_{o}\rangle$.} \label{modifiedCNOT}
\end{figure}

Fortunately, however, it is possible to construct a photonic circuit
in which the phase shifts introduced by dispersion can be equalized.
A Ti:LiNbO$_3$ photonic quantum circuit that behaves as a novel
dispersion-managed, deterministic, two-qubit, single-photon, CNOT
gate is sketched in Fig.~\ref{modifiedCNOT}. It makes use of three
paths (upper, middle, and lower), in which the path-lengths of the
three arms are carefully adjusted to allow for dispersion
management. The third path provides the additional degree-of-freedom
that enables the optical path-lengths to be equalized.

The design relies on the use of polarization-dependent mode
analyzers at the input to the circuit. The TM-mode analyzer couples
the odd-TM component of the state to the upper path, while the
TE-mode analyzer couples the odd-TE component to the lower path. The
even-TM and even-TE components continue along the middle path.
Polarization-dependent mode combiners are used at the output of the
circuit.

If the control qubit is in a superposition state, the general
quantum state at the input to the circuit, which resides in a 4D
Hilbert space (2D for polarization and 2D for mode number), is
expressed as
\begin{equation}
\begin{array}{rl} \label{eq:PsiIn}
\vert\Psi_{i}\rangle= &\alpha_{\,1}\vert e,\mathrm{TM}\rangle+\alpha_{\,2}\vert o,\mathrm{TM}\rangle+ \alpha_{\,3}\vert e,\mathrm{TE}\rangle+\alpha_{\,4}\vert o,\mathrm{TE}\rangle \vspace{1mm} \\
=& \vert \mathrm{TM}\rangle\otimes\left[\,\alpha_{\,1}\vert
e\rangle+\alpha_{\,2}\vert o\rangle\,\right] +\vert
\mathrm{TE}\rangle\otimes\left[\,\alpha_{\,3}\vert
e\rangle+\alpha_{\,4}\vert
 o\rangle\,\right] \vspace{1mm} \\
=&\vert e\rangle\otimes\left[\, \alpha_{\,1}\vert
\mathrm{TM}\rangle+\alpha_{\,3}\vert \mathrm{TE}\rangle\,\right]
+\vert o\rangle\otimes\left[\,\alpha_{\,2}\vert
\mathrm{TM}\rangle+\alpha_{\,4}\vert \mathrm{TE}\rangle\,\right],
\end{array}
\end{equation}
where $|e\rangle$ and $|o\rangle$ are the basis states of the modal
subspace; $|{\rm TM}\rangle$ and $|{\rm TE}\rangle$ are the basis
states of the polarization subspace; the $\alpha$'s represent the
basis weights; and $\otimes$ indicates the tensor product. Since the
target (modal) qubit is flipped by a TM control qubit, the output
state $\vert\Psi_{o}\rangle$ becomes
\begin{equation}\begin{array}{rl}\label{eq:PsiOut}
 \vert\Psi_{o}\rangle=&\alpha_{\,1}\vert o,\mathrm{TM}\rangle+\alpha_{\,2}\vert e,\mathrm{TM}\rangle+ \alpha_{\,3}\vert e,\mathrm{TE}\rangle+\alpha_{\,4}\vert o,\mathrm{TE}\rangle \vspace{1mm}  \\
 =& \vert \mathrm{TM}\rangle\otimes\left[\,\alpha_{\,1}\vert o\rangle+\alpha_{\,2}\vert e\rangle\,\right] +\vert \mathrm{TE}\rangle\otimes\left[\,\alpha_{\,3}\vert e\rangle+\alpha_{\,4}\vert
 o\rangle\,\right],
\end{array}
\end{equation}
where it is clear that the two terms in the input state,
$\alpha_{\,1}\vert e,\mathrm{TM}\rangle$ and $\alpha_{\,2}\vert
o,\mathrm{TM}\rangle$, are converted to $\alpha_{\,1}\vert
o,\mathrm{TM}\rangle$ and $\alpha_{\,2}\vert e,\mathrm{TM}\rangle$,
respectively, at the output, exemplifying the operation of this CNOT
gate. Figure~\ref{modifiedCNOT} displays the paths taken by the
components of the input state provided in Eq.~(\ref{eq:PsiIn}); the
output state set forth in Eq.~(\ref{eq:PsiOut}) is also indicated.

The output state in Eq.~(\ref{eq:PsiOut}) is entangled in polarization and mode number; it is inseparable and cannot be written in factorizable form. A particular property of the CNOT gate is the induction of entanglement between factorized qubits: if the control qubit is in the superposition state $\frac{1}{\sqrt2}[\,|{\rm TM}\rangle + |{\rm TE}\rangle]$, and the target qubit is in one of the computational basis states, then the output state of the CNOT gate is maximally entangled. An experimental test of the entanglement created between the polarization and modal degrees-of-freedom can be effected by using quantum-state tomography. The input to the CNOT gate can be readily generated from a product state, say $|{\rm TM}\rangle \otimes |e\rangle $, by rotation using a waveguide-based EO TE$\rightleftharpoons$TM mode converter \cite{Yariv73,Alferness80}, in addition to a phase modulator, as described in Sec.~6.

It remains to demonstrate the manner in which dispersion management
can be achieved in the CNOT gate displayed in
Fig.~\ref{modifiedCNOT}. The phase shift $\varphi$ acquired by each
component at the output is given by
\begin{equation}
\begin{array}{cl}
\varphi_{\,\,e,{\rm TM}}& =\beta_{\,e,{\rm TM}}
\,\ell_{1}+\beta_{o,{\rm TM}}\, \ell_{2}+\beta^{\prime}
\,L_{D}-\left( 2q_{1}+q_{2}\right) \pi/2
\\[1mm]
\varphi_{\,\,o,{\rm TM}}& = \varphi_{\,\,e,{\rm TM}}\\[1mm]
\varphi_{\,\,e,{\rm TE}} &= 2\beta_{\,e,{\rm TE}} \,\ell_{1}+\beta^{\prime\prime}\, L_{D}\\[1mm]
\varphi_{\,\,o,{\rm TE}}& =2\beta_{\,o,{\rm TE}}\, \ell_{3}
-q_{3}\pi +2\phi_{A}\,,
\end{array}
\end{equation}
where the $\beta$'s are the mode propagation constants;
$\beta^{\prime}$ is the propagation constant of either the TM-even
mode in WG1 or the TM-odd mode in WG2; $\beta^{\prime\prime}$ is the
propagation constant of the TE-even mode in WG1; $q_{1}$, $q_{2}$,
and $q_{3}$ are odd positive integers that depend on the lengths of
the TM-mode analyzer, directional coupler, and TE-mode analyzer,
respectively; $L_{D}$ is the length of the directional-coupler
electrode; $\ell_1$ is the path-length for the even modes before and
after the directional coupler, $\ell_2$ is the path-length for the
odd-TM mode before and after the directional coupler; and $2\ell_{1}
+L_{D}$, $2 \ell_{2}+L_{D}$, and $2\ell_{3}$ are the overall
physical lengths of the middle, upper, and lower paths,
respectively. The phase shift $\phi_{A}$ arises from the coupling
that affects the odd-TE component as it travels through the TM-mode
analyzer. Phase shifts that accrue for the even modes as they pass
through the mode analyzers and mode combiners are neglected because
of large phase mismatches and weak coupling coefficients. By
adjusting the lengths $\ell_{1}$, $\ell_{2}$, and $\ell_{3}$, we can
equalize the phase shifts encountered by each component of the
state. Imperfections in the fabrication of the circuit may be
compensated by making use of EO phase modulators.

\begin{figure}
\centering
\includegraphics[width=5 in,totalheight =2.8 in]{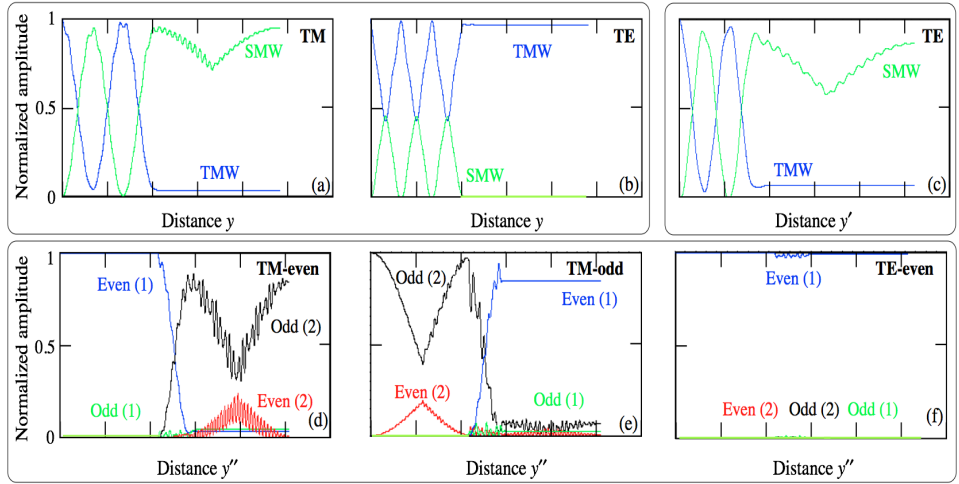}
\caption{Simulation demonstrating the performance of the
polarization-dependent mode analyzers and the EO TMW directional
coupler associated with the dispersion-managed, deterministic,
two-qubit, single-photon, CNOT gate set forth in
Fig.~\ref{modifiedCNOT}. The input wavelength is $\lambda = 0.812
\:\mu$m. The TM-mode-analyzer and mode-combiner parameters are
$w_{1}= 5.6 \:\mu$m, $w_{2}=3.4\:\mu$m, $b_1=4\:\mu$m, and $L_{2}=
6.2$ mm; the TE-mode-analyzer and mode-combiner parameters are
$w_{2}=3 \:\mu$m, $b_1=4 \:\mu$m, and $L_{2}= 3.7$ mm (see
Fig.~\ref{Modeanalyzer} for symbol definitions). The $S$-bends have
dimensions $L_{b} = 10$ mm and $S = 127 \:\mu$m. The TMW
directional-coupler has length $L_1 = 2.2$ mm, waveguide width
$w_{1}= 5.6 \: \mu$m, electrode separation $d = 4\:\mu$m, and an EO
phase-modulator voltage $V= 36$ V applied to WG2, with WG1 at ground
potential. All panels display the spatial evolution of the
normalized amplitudes of the interacting modes. (a) The curves
display strong coupling between the odd and even modes for
TM-polarization inside the TM-mode analyzer. The input odd mode in
the TMW is shown in blue and the even mode transferred to the SMW is
shown in green [the same color conventions are used in panels (b)
and (c)]. The even mode is ultimately coupled to another TMW at the
output of the TM-mode analyzer and once again becomes odd. (b) The
curves show negligible coupling between the odd and even modes for
TE-polarization inside the TM-mode analyzer. (c) The curves display
good coupling between the odd and even modes for TE-polarization
inside the TE-mode analyzer. At the TE-mode combiner, the even mode
in the SMW once again becomes an odd mode in the TMW. Panels (d),
(e), and (f) display the performance of the directional coupler for
modal inputs that are TM-even, TM-odd, and TE-even, respectively.
For a given polarization, the  blue and green curves represent the
amplitudes of the even [denoted Even(1)] and odd [denoted Odd(1)]
modes in WG1, respectively, while the the red and black curves are
the amplitudes of the even [denoted Even(2)] and odd [denoted
Odd(2)] modes in WG2, respectively.
All simulated data in this figure were obtained using the RSoft
program.}\label{ExCNOT}
\end{figure}
A simulation that demonstrates the performance of the
polarization-dependent mode analyzers and EO TMW directional coupler
is presented in Fig.~\ref{ExCNOT}. The lengths $\ell_{1}$,
$\ell_{2}$, and $\ell_{3}$ are assumed to be adjusted such that they
equalize the phase shifts encountered by each component of the state
so that dispersion is not an issue. The spatial evolution of the
normalized amplitudes of the odd and even modes inside the TM-mode
analyzer, for TM- and TE-polarization, are displayed in
Figs.~\ref{ExCNOT}(a) and \ref{ExCNOT}(b), respectively. It is
apparent that the TM-mode analyzer extracts only the TM-odd
component, while the TE-odd component remains in the TMW waveguide
until it couples to the lower path via the TE-mode analyzer [see
Fig.~\ref{ExCNOT}(c)]. Figures~\ref{ExCNOT}(d), (e), and (f) display
the performance of the directional coupler for modal inputs that are
TM-even, TM-odd, and TE-even, respectively.
It is apparent in Fig.~\ref{ExCNOT}(d) that the power in the even
mode in WG1 is transferred to the odd mode in WG2 for TM
polarization. Figure~\ref{ExCNOT}(e) reveals complementary behavior:
the power in the odd mode in WG2 is transferred to the even mode in
WG1. Figure~\ref{ExCNOT}(f), on the other hand, shows that the
TE-even mode travels through the directional coupler with
essentially no interaction. Figures~\ref{ExCNOT}(d), (e), and (f),
taken together, along with the observation that the TE-odd mode
preserves its modal profile during propagation, demonstrate a flip
of the modal target qubit by the TM-polarized control qubit, and no
flip by a TE-polarized control qubit, confirming that the photonic
circuit in Fig.~\ref{modifiedCNOT} behaves as a CNOT gate.

The absence of a total power transfer from one waveguide to another
in Figs.~\ref{ExCNOT}(d) and \ref{ExCNOT}(e) can be ascribed to
sub-optimal simulation parameters. The conversion efficiency can be
expected to improve upon: 1) optimizing the length of the two-mode
directional coupler; 2) minimizing bending losses by increasing the
length of the $S$-bend; 3) mitigating the residual phase mismatch by
more careful adjustment of the voltage; and 4) improving numerical
accuracy. Moreover, the deleterious effects of dc drift and
temperature on the operating voltage and stability of the two-mode
directional coupler can be minimized by biasing it via electronic
feedback \cite{Djupsjobacka89}; a novel technique based on inverting
the domain of one of its arms can also be used to reduce the
required operating voltage \cite{Lucchi07}. Finally, it is worthy of
note that decoherence associated with the use of a cascade of CNOT
gates, such as might be encountered in carrying out certain quantum
algorithms, may be mitigated by the use of either a qubit amplifier
\cite{gisin10} or teleportation and error-correcting techniques
\cite{Glancy04}.

\section{Conclusion}
The modes of a single photon in a two-mode Ti:LiNbO$_{3}$ waveguide
have been co-opted as basis states for representing the quantum
state of the photon as a modal qubit. Various photonic quantum
circuit designs have been presented for carrying out basic
operations on modal qubits for quantum information processing
applications. These include a mode analyzer, a mode rotator, and
modal Pauli spin operators. We have also described the design of a
deterministic, two-qubit, single-photon, CNOT gate, as well as a
dispersion-managed version thereof, that rely on a single photon
with both modal and polarization degrees-of-freedom in a joint 4D
Hilbert space. The CNOT gate is a key element in certain sets of
universal quantum logic gates. Simulations of the performance of all
of these components, carried out with the help of the the commercial
photonic and network design software package RSoft, provide support
that they operate as intended. The design of these devices is based
on selective and controllable power coupling among waveguides, the
isomorphism between waveguide coupling and the SO(2) rotation
matrix, and the tensor polarization properties of the Pockels
coefficients in lithium niobate. The flexibility of Ti:LiNbO$_{3}$
as a material for the fabrication guided-wave structures should
accommodate the development of increasingly complex quantum circuits
and serve to foster new architectures.

\section*{Acknowledgments}
This work was supported by the Bernard M. Gordon Center for
Subsurface Sensing and Imaging Systems (CenSSIS), an NSF Engineering
Research Center; by a U.S. Army Research Office (ARO)
Multidisciplinary University Research Initiative (MURI) Grant; and
by the Boston University Photonics Center.

\end{document}